\documentclass{ecai} 

\usepackage{caption}
\usepackage{subcaption}
\usepackage{float}
\usepackage{hyperref}
\usepackage{array}
\newcolumntype{C}[1]{>{\centering\let\newline\\\arraybackslash\hspace{0pt}}m{#1}}
\usepackage{graphicx}
\usepackage{multirow}
\usepackage{siunitx}
\usepackage{subcaption}
\usepackage{graphicx}
\usepackage{svg}
\usepackage{booktabs}
\usepackage{xcolor}
\usepackage{textcomp}
\usepackage{xcolor}
\usepackage{soul}
\usepackage{tikz}
\usepackage{algorithm}
\usepackage{float} 
\usepackage{array}
\usepackage[noend]{algpseudocode}

\algdef{SE}[DOWHILE]{Do}{doWhile}{\algorithmicdo}[1]{\algorithmicwhile\ #1}%
\usetikzlibrary{arrows}

\newcommand{\BibTeX}{B\kern-.05em{\sc i\kern-.025em b}\kern-.08em\TeX}

\begin{document}


\begin{frontmatter}


\paperid{123} 


\title{Graphite: A Graph-based Extreme Multi-Label Short Text Classifier for Keyphrase Recommendation}


\author[A]{\fnms{Ashirbad}~\snm{Mishra}
}
\author[B]{\fnms{Soumik}~\snm{Dey}}
\author[B]{\fnms{Jinyu}~\snm{Zhao}} 
\author[B]{\fnms{Marshall}~\snm{Wu}}
\author[B]{\fnms{Binbin}~\snm{Li}}
\author[A]{\fnms{Kamesh}~\snm{Madduri}}

\address[A]{The Pennsylvania State University}
\address[B]{eBay Inc.}


\begin{abstract}
 \textit{Keyphrase Recommendation} has been a pivotal problem in advertising and e-commerce where advertisers/sellers are recommended keyphrases (search queries) to bid on to increase their sales. It is a challenging task due to the plethora of items shown on online platforms and various possible queries that users search while showing varying interest in the displayed items. Moreover, query/keyphrase recommendations need to be made in real-time and in a resource-constrained environment. This problem can be framed as an Extreme Multi-label (XML) Short text classification by tagging the input text with keywords as labels. Traditional neural network models are either infeasible or have slower inference latency due to large label spaces. We present \emph{Graphite}, a graph-based classifier model that provides real-time keyphrase recommendations that are on par with standard text classification models. Furthermore, \textit{it doesn't utilize GPU resources}, which can be limited in production environments. Due to its \textit{lightweight nature} and \textit{smaller footprint}, it can train on very large datasets, where state-of-the-art XML models fail due to extreme resource requirements. Graphite is deterministic, transparent, and intrinsically more interpretable than neural network-based models. We present a comprehensive analysis of our model's performance across forty categories spanning eBay's English-speaking sites.
\end{abstract}

\end{frontmatter}


\section{Introduction}
\label{s:intro}
In the online search space, the sponsored search mechanism~\cite{jansen2008sponsored} promotes paid entities to be shown above or beside the standard search results retrieved for a user query. The promoted entity can be shown as an advertisement to the web user or as an e-commerce listing to a buyer corresponding to their search queries. The search engine platform facilitates advertiser/seller \emph{bidding} on search queries. The recommendation engine automatically ranks relevant search queries given a single or a group of e-commerge listings. Advertisers or sellers bid via large-scale \emph{campaigns} to increase the visibility of their products. The solution to the recommendation task will be advantageous to both small and large-scale advertisers/sellers increasing their sales and the platform's revenues. 

Query (or keyphrase) recommendations can easily be enabled from short texts representing the entity being sold/advertised, such as titles of e-commerce products, product reviews, transcripts of promotional videos, posts on social media sites, and other search-based platforms using sponsored entities~\cite{bartz2006logistic,yih2006finding}. Here, we focus on e-commerce platforms such as Amazon, eBay, and Walmart, where the sellers are recommended keyphrases based on their inventory listing's metadata. Domain experts have found that a listing/item's title is paramount as they are quite specific, and other metadata can be missing or incoherent. For instance, a seller may have a listing titled \emph{``New iPhone 15 Pro Max 128 GB White"} in the Electronics category. We seek to develop a recommender algorithm that will generate a relevant collection of keyphrases such as \textit{``latest iphone"}, \textit{``128 gb iphone 15"}, \textit{``new iphone"}, \textit{``apple phone 15"}, or even \textit{``latest Samsung smartphone''}. We use \emph{keyphrase} in this paper instead of the more common \emph{keyword} to indicate that a phrase can have multiple words; also, the ordering of words in the phrase matters. The listing's length in words is typically very low, and the keyphrase may contain words that do not even occur in the listing. 

Conventionally, logs generated by search engine responses \cite{aglo2000QueryLog,bartz2006logistic} to buyer queries have been the data source for recommendation systems. This helps to associate keyphrases that are not only relevant to the products, but are also actively searched by buyers. We have provided more details on the data generation process in Section~\ref{sss:eBay_datasets}. Aggregated data from the logs typically contains billions of data points due to the large number of \emph{tail keyphrases} (queries searched less often). Moreover, relative to the size of the data, there are limited resources for the execution of recommendation systems. Scenarios such as the new setup of a seller's inventory require real-time or near-real-time recommendation. Thus, effective strategies should be able to suggest a small subset of relevant keyphrases from a large space of keyphrases in a constrained environment with suggestions provided in real-time.

Advertisers only want to bid on keyphrases that are actual queries and not queries that seem plausible but non-existent for targeting purposes. Since the nature of the problem is mapping items to \emph{multiple} queries to increase the potential reach of the advertisers, this problem of keyphrase recommendation can be formulated as an Extreme Multi-Label (XML) Classification problem. Keyphrase recommendation has been explored using query-query similarity with click graphs~\cite{antonellis2008simrank++}, and formulated as an XML problem in~\cite{agrawal2013multi}.

Typically, the keyphrase-recommendation datasets in the advertisement domain exhibit a \emph{power-law distribution}, with a large number of tail keyphrases. Therefore, it is difficult to cluster words that occur within listings in a category, diminishing the effectiveness of many text classifiers. On the other hand, the text classifiers that provide effective predictions have high inference cost, which is usually the issue with Large Language Models (LLMs)~\cite{kaddour2023challenges} such as BERT~\cite{devlin-etal-2019-bert}, GPT-3~\cite{brown2020language}, etc.

The datasets under consideration are provided by eBay from their proprietary search logs. The ``Very large'' categories in the dataset have more than two million of keyphrases associated with more than three million training data points. Hence, the coverage of keyphrases in the training data is extremely sparse. With more products being added to the platform every second along with buyers constantly querying for new products, the datasets are expanding at an accelerated pace. To match this increase, models need to exhibit a low memory footprint and faster execution in both training and inference, while also scaling well according to the number of labels and training data points.  

Our contributions and scope in this work are as follows:
\begin{itemize}
    \item We develop a novel bipartite graph-based model that is simple and interpretable for commercial purposes.
    \item Our developed model is the most efficient among all the state-of-the-art (SOTA) models for keyphrase/query recommendation in the sponsored search domain with real-time inferencing. It has a low memory footprint, extremely low training time, and parallelized inferencing that scales efficiently and reliably for very large datasets with millions of labels. Also, under resource-constrained settings, it achieves the highest accuracy among the SOTA models.
    \item We give performance comparisons on real-world search-related datasets from the eBay e-commerce platform and public datasets. Our model handles various practical hurdles such as the \emph{cold start} issue, optimized resource usage for real-time utility, and integration complexity with existing pipelines.
    \item We show some commercial impacts of the model in real-world scenarios.
\end{itemize}




\section{Related Work}
\label{sec:rel_work}

Bipartite graphs have long been used in numerous domains to model user search queries from the logs such as query-url graphs \cite{aglo2000QueryLog,baeza2007extracting} and query-ad graphs \cite{anastasakos2009collaborative}. Generally, methods operating on these bipartite graphs compute similarities between queries based on the items they are associated with, which are then used to recommend the queries for new items. Simrank++ \cite{antonellis2008simrank++} improves the query similarity technique by decreasing the number of iterations needed for convergence. In addition, it improves the generated score by multiplying them by a factor depending on the number of common neighbors of those queries. However, in the worst case, such approaches would require a comparison between each pair of queries (i.e. quadratic scaling), which is infeasible when there are a large number of keyphrases. In addition, the ranking of recommended queries associated with similar items based on the relevance to the item is an issue.

The state-of-the-art models for XML problems~\cite{dahiya2023ngame,renee_2023,Dahiya23b,you2019attentionxml,dahiya2021deepxml} use deep neural networks (DNNs), typically with one-vs-all (OVA) classifiers. While the models~\cite{dahiya2023ngame,Dahiya23b} require label features to work with, \emph{AttentionXML}~\cite{you2019attentionxml}, \emph{Renee}~\cite{renee_2023} and \emph{DeepXML/Astec}~\cite{dahiya2021deepxml} can work without them. Among the DNN models in the benchmarked studies~\cite{Bhatia16}, we find that the DeepXML/Astec model~\cite{dahiya2021deepxml} is able to scale to large datasets (e.g., the public AmazonTitles-3M dataset) and has lower training time compared to competing methods. Also, the authors of~\cite{dahiya2021deepxml} show that Astec achieves real-time inference latencies. Therefore, given the scope of this work, we find Astec to be the most suitable for comparison.

DeepXML/Astec \cite{dahiya2021deepxml} is a pipelined framework that processes the task of classifying texts end-to-end. Each component of the framework can be replaced to mimic different classification algorithms. Unlike another extreme classifier Slice \cite{jain2019slice}, Astec generates its own embedding, which is otherwise expensive to compute. In terms of implementation and training, DeepXML consists of 3 stages. The first stage trains a \emph{surrogate task} that generates an intermediate representation. Faster training is enabled by reducing the label space, which is done by first generating label representations from the representations of input texts (instance) that the label is associated with in the training set. Clusters are generated by associating together labels with similar representations and annotating each cluster as a meta-label. The second stage (called \emph{extreme}) trains OVA classifiers for each label and optimizes training by \emph{shortlisting} labels for each data point using negative sampling. The third stage called \emph{reranker}, is similar to the \emph{extreme} stage, which uses the pre-trained shortlists from the extreme stage.

The fastText~\cite{fastText1,fastText2} software tool/model has proven to be a CPU-based, efficient solution for handling large workloads. fastText generates word vectors using the \emph{CBOW} model and uses a simple linear neural network model with hierarchical softmax for faster training and inference. One of the reasons why fastText works well is because it incorporates \emph{subword} information into its embeddings. The size of the model can easily be constrained to use less storage space using techniques such as quantization~\cite{fasttext3}, and pruning the vocabularies of keyphrases and title words.

fastText is \emph{one of the models deployed at eBay} for real-time keyphrase recommendation to sellers. Hence, we use it as a baseline for our comparison.


\section{Graphite Model}
\label{s:graphite}
In this section, we introduce some notation and formally define the problem we are trying to solve. Subsequently, we describe the two essential steps of our model: the \emph{Construction} step equivalent to training, and the \emph{Inference} step. And then discuss the implementation details.
\vspace{-3mm}
\subsection{Problem Formulation and Notations}
\label{ss:notations}
We formally define the \textit{Multi-label Text Classification} problem where there are multiple labels associated with each instance (input text). Each instance consists of a list of words from the textual data of the instance and a list of labels that will be modeled by a classifier. Such a model is generally constructed or learned from the training part of the dataset $T(A,B)$, which consists of two sets of lists $A$ and $B$ of equal size, representing instances and labels respectively. Each $a_i\in A$ is a list $a_i=\{w_1,w_2,...,\}$, where each $w_*$ is a word in the instance $a_i$. Similarly, each list $b_i\in B$ has the form $b_i=\{l_1,l_2,...\}$, where each $l_*=\{w_1,w_2,...\}$ is a label consisting of a list of words. Together, $a_i$ and $b_i$ denote the training sample for index $i$. Note that the ordering of elements in the label list is essential and list permutations create unique labels. Formally, a Bipartite Graph $G(V,E)$ has two disjoint subsets of vertices $X$ and $Y$ such that $X\bigcup Y=V$ and $X\bigcap Y=\phi$. The edges of the set $E$ connect a vertex in $X$ with a vertex in $Y$ and there are no edges connecting the vertices within each subset. We define the function \textit{Search, Deduplicate and Count} or $SDC(\cdot)$ which given a list of elements, counts the occurrences of each element in the list. It outputs a list of tuples of the form $(element,count)$ for each unique element in the list. We also define functions \textit{Create Map} or $cmap(x,y)$ that maps $x$ to $y$ and \textit{Get Mapping} or $gmap(x)$ that retrieves the mapping for $x$. $gmap(*)$ returns all the mapping as a list. In XML systems the instance is the item/document in question and the labels are the keyphrases. Graphite's main idea is that two items should be associated with the same keyphrase if they are similar. For inference, given a test item, our model identifies items from the training data that are similar to the test item. The labels associated with the similar items are then ranked in order of relevance to the test item. 


\subsection{Model Construction}
\label{ss:graphite_const}
The Graphite model constructs two bipartite graphs, $G_{WI}(V,E)$ and $G_{IL}(V,E)$ from the training set $T(A,B)$. $G_{WI}(V,E)$ maps the words within each instance/sample to the identifier (or index) of the instance/sample. Identifiers are nonnegative integer numbers of instances in the set $T$. The vertex set $V\in G_{WI}$ consisting of two disjoint sets, $X$ and $Y$, corresponding to the unique set of words in the instance and unique instance identifiers, respectively. So, $X=\bigcup_{\forall w\in a_i,\forall a_i\in A} \{w\},A\in T$ and $Y=\{0,1,...,|A|\}$, where $|\cdot|$ denotes cardinality. The edges $E\in G_{WI}$ are constructed from the tuples as $E=\{(w,i), \forall w\in a_i\in A\}$. Next, our model constructs the graph $G_{IL}(V,E)$ that maps the instance identifiers to their associated labels in $T$. Similarly, the disjoint subsets $X$ and $Y$ of the vertex set $V\in G_{IL}$ are built as follows. $X=\{0,1,...,|A|\}$ and $Y=\bigcup_{\forall l\in b_i,\forall b_i\in B} \{l\},B\in T$ where each $l$ is a label. The edge set $E\in G_{IL}$ is the unique set of tuples $E=\{(i,l), \forall l\in b_i\in B\}$. Note that the ordering of the instance identifiers does not matter.

\begin{figure}[h]
    \scriptsize
    \centering
    \begin{subfigure}[b]{\linewidth}
        \centering
        \begin{tabular}{@{}l|l|l@{}}
        \toprule
        \textbf{ID} & \textbf{Keyphrases}    &    \textbf{Item Titles}   \\ \midrule
        1  & \begin{tabular}[c]{@{}l@{}}iphone 12 pro, black phone\end{tabular} & black iphone 12 pro 128GB \\
        2  & pixel 6, black phone                             &    google pixel black 64GB     \\
        3  & iphone 13 pro, grey phone                         & grey iphone 13 pro  \\
        4  & Samsung galaxy, grey phone                              & Samsung s6 grey \\ \bottomrule
        \end{tabular}
        \caption{Illustrated Training Data}
    \end{subfigure}
    \par\bigskip 
    \begin{subfigure}[b]{\linewidth}
        \centering
        \includegraphics[width=0.4\linewidth]{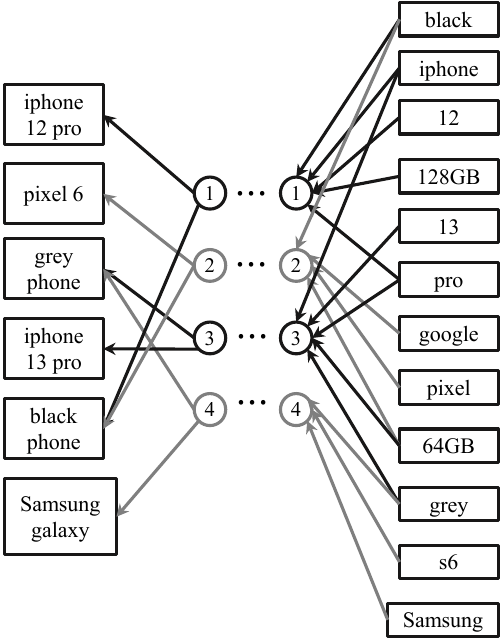}
        \par\bigskip 
        \caption{Tripartite Graph derived from Illustrated Data}
    \end{subfigure}
    \par\bigskip 
    \caption{Illustration of Graphite's construction phase. Subfigure (a) shows a set of item's titles along with their associated keyphrase and subfigure (b) shows the graphs $G_{WI}$ and $G_{IL}$ that are constructed from the set in (a).}
    \par \medskip
    \label{fig:graphite_const_eg}
\end{figure}

Together, both graphs $G_{WI}$ and $G_{IL}$ map the words in an item to the associated keyphrases akin to a tripartite graph. Modelling a tripartite graph as two bipartite graphs helps us construct and store the data efficiently. We show an illustration of the structure and contents of the graphs in figure~\ref{fig:graphite_const_eg}. Subfigure (a) displays a set of sample training instances as the item's title along with the associated labels as keyphrases. Subfigure (b) shows the graphs $G_{WI}$ and $G_{IL}$ constructed from the sample training set using Graphite's construction phase. The graph $G_{WI}$ is constructed by mapping the words in the title to the items ids (training instance ids), and the graph $G_{IL}$ is constructed by mapping those items ids to the keyphrases. The visual shows intuitively how the keyphrases can be mapped from the words, thus, for a test instance determining relevant keyphrases is quite evident.
\vspace{-3mm}

\subsection{Inference Step}
\label{ss:graphite_inf}
Given a test instance $t$ with the words in the title, the classifier predicts a list of labels in order of relevance to the instance. Graphite's inference step executes two phases \emph{Clustering} followed by \emph{Ranking} on the test instance $t$ which is described in the subsections below. The goal of the Clustering phase is to cluster candidate labels into groups based on their similarity to the test instance. Each candidate label is obtained from the training instances represented in $G_{WI}$ and $G_{IL}$. The Ranking step re-ranks the labels in each group/cluster based on label attributes generated during the clustering phase.

\begin{algorithm}[h]
\caption{Graphite's Inference}

\begin{algorithmic}[1]
\scriptsize
\Require Graphs $G_{WI}$ and $G_{IL}$ and test instance $t$
\Ensure List of lists ($R$) with labels and their attributes
\Function{Clustering}{$G_{WI}$,$G_{IL}$,$t$}
\State $I,L,R \gets []$ \Comment{Lists of instances, labels and results resp.}
\For {$w$ in $t$}
    \For {$(w,i)$ in $E\in G_{WI}$}
        \State $I\gets I+i$
    \EndFor
\EndFor
\State $I\gets SDC(I)$
\For {$(i,c)$ in $I$}
    \For {$(i,l)$ in $E\in G_{IL}$}
        \State $L\gets L+l$
        \State $cmap(l,c)$
    \EndFor
\EndFor
\State $L\gets SDC(L)$
\For{$i\gets gmax(map(*))$ to $1$}
    \State $C\gets[]$
    \For{$(l,m)$ in $L$ and $gmap(l)==i$}
        \State{$C\gets C+(l,WMR(t,l),m)$}
    \EndFor
    \State $R\gets R+C$
\EndFor
\State \textbf{return} $R$
\EndFunction
\end{algorithmic}

\label{alg:graphite_cluster}
\end{algorithm}
\vspace{-6mm}

\subsubsection{Clustering Phase}
The algorithm~\ref{alg:graphite_cluster} describes the clustering phase. Given a test instance $t$, it generates clusters of the candidate labels as a list. Each generated candidate label is then associated with a set of attributes required for the next ranking step. The algorithm first starts by mapping the words in the test instance $t$ to the list of instance ids ($i$) in $G_{WI}$ as shown in lines 3-5 of algorithm~\ref{alg:graphite_cluster}. The $SDC(\cdot)$ function\footnote{\label{fn:terms}Defined in section~\ref{ss:notations}} is utilized in lines 6 and 11. The output ($I$) of line 6 is a list of tuples with instance IDs ($i$) and the occurrence count ($c$) of the instance. The occurrence count is termed as \emph{similarity (score)} as it is equivalent to the number of similar words to the test instance $t$. Next, lines 7-10 use $G_{IL}$ to find the labels associated with the instances in $I$. The similarity score ($c$) of each instance ($i$) is assigned to the label ($l$) associated with it using the $cmap(\cdot,\cdot)$ function\footref{fn:terms}. Labels with two different similarity scores are mapped to the higher score. The $gmap(\cdot)$ function\footref{fn:terms} returns the similarity score of a label and $gmap(*)$\footref{fn:terms} returns all similarity scores. After the execution of line 11, the list $L$ contains candidate labels with their \emph{Multiplicity} ($m$) which indicates the number of unique instances from which the label was derived. The for loop in line 12 iterates from the largest similarity score to the smallest, to create a list of clusters $R$. Inner loop 14 ensures that each cluster only groups the labels with the same similarity score. Each element of the cluster is a tuple containing the label $l$ and its two attributes, \emph{Word Match Ratio} computed by the function $WMR(t,l)=\frac{|t\cap l|}{|l|}$ and \emph{Label Multiplicity} given by $m$.

\subsubsection{Ranking Phase}
\label{sss:graphite_ranking}
The ranking phase operates on the list of clusters $R$ obtained at the end of the algorithm~\ref{alg:graphite_cluster}. The clusters in $R$ are in non-increasing order of the similarity score, i.e, each label in the first cluster has the highest similarity score followed by labels with second highest similarity score and so on. This partial ordering provides us with some relevancy ranking, but the labels within each cluster are not ranked. Thus, during this phase, the labels within each cluster are ranked using their attributes. The labels are ranked in the non-increasing order of \emph{word match ratio} and to break the ties the label with larger \emph{label multiplicity} is ranked first. Our model follows a \emph{label word-aware} technique that looks at the similarity between the label and the test instance. This is enabled by the \emph{word match ratio} where labels with more common words with test instance are preferred. Note that the denominator in the function $WMR(\cdot,\cdot)$ ensures that smaller-length labels are preferred. We find that this works best for the dataset under consideration, while it can be tweaked for other datasets. \emph{Multiplicity} of a label indicates how many similar instances were associated with the label, indicating a higher probability of relevance.

\subsection{Implementation Details}
\label{sss:illustration_implementation}

The construction of both $G_{WI}$ and $G_{IL}$ is first done by building a list of tuples indicating the edges, then edges are sorted, de-duplicated and stored in \emph{Compressed Sparse Row (CSR)} format. The time complexity is log-linear and the space complexity is linear in the number of edges. The number of edges is asymptotically $O(|A|\cdot \max|a_*|)$ for $G_{WI}$ and $O(|B|\cdot \max|b_*|)$ for $G_{IL}$. In the implementation, the words and the labels are represented as \textit{unsigned integers} to reduce storage costs and avoid string comparisons. In other words, comparing two words or two labels takes constant time. The training step does not involve any weight updates or hyper-parameter training, making it quite fast and efficient.

Generally, during inference, a predetermined number of labels is required to be predicted. Due to the large label spaces, our method would output a large number of predictions, many of which would be irrelevant. Hence, we limit the number of predictions by only considering the clusters in $R$ with the highest similarity. We preemptively achieve this by modifying the function $SDC(\cdot)$ in 
Algorithm~\ref{alg:graphite_cluster} line 6 to only include a limited number of instances in $I$ as required by the number of predictions. Instances with higher similarity scores are only picked while ensuring that if a unique similarity score is considered then all instances with that similarity score are picked. Thus, at the end of the ranking phase, only the required number of ordered labels are returned. The list of clusters $R$ is implemented by extending the length of each tuple to include the similarity score. Thus, the ranking phase first orders $R$ by the similarity scores of each label and then breaks ties using both the label attributes. 

\begin{figure*}[t]
    \centering
    \scriptsize
     \begin{subfigure}[b]{0.8\linewidth}
         \centering
         \includegraphics[width=\linewidth]{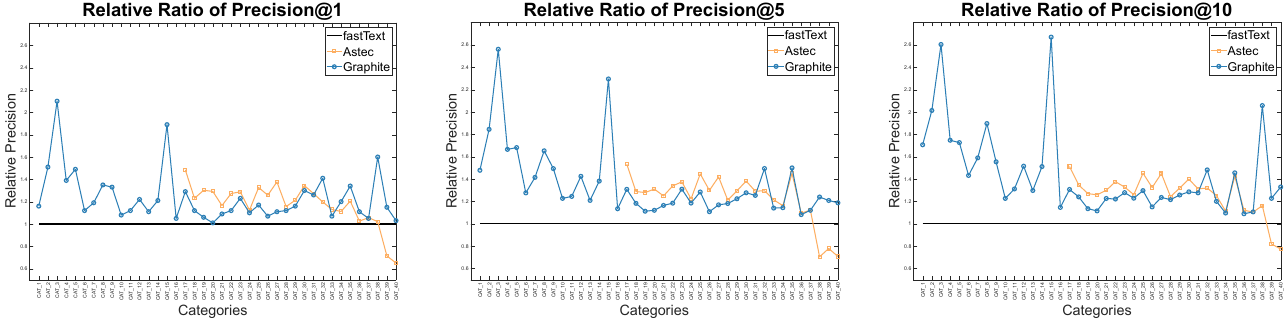}
         \caption{Precision@1/5/10 ratios of fastText, Astec and Graphite}
         \label{fig:graphite_scores_precision}
    \end{subfigure}
    \par\bigskip 
     \begin{subfigure}[b]{0.8\linewidth}
         \centering
         \includegraphics[width=\linewidth]{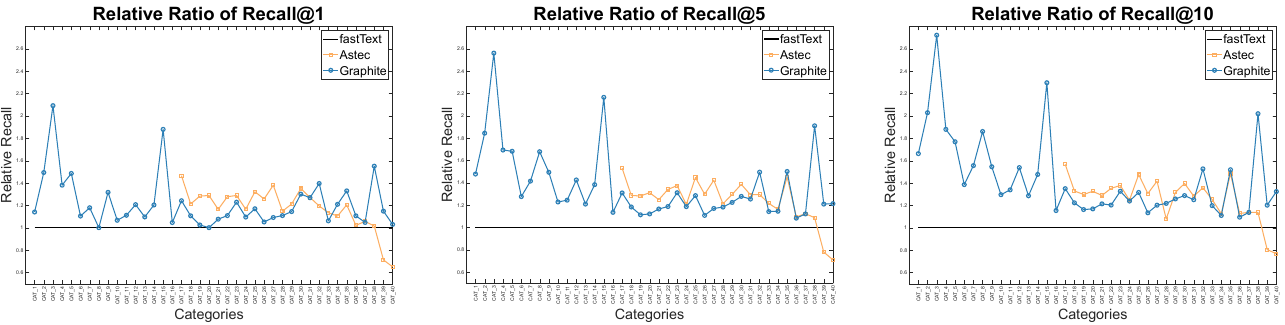}
         \caption{Recall@1/5/10 ratios of fastText, Astec and Graphite}
         \label{fig:graphite_scores_recall}
    \end{subfigure}
    \par\bigskip 
    \caption{Comparison of Precision and Recall scores of fastText, Astec and Graphite with top $10$ predictions from each model.}
    \label{fig:graphite_scores}
    \par \medskip
\end{figure*}

The functions $SDC$, $cmap$ and $gmap$ are implemented using log-linear time data structures. The ranking phase sorts the list of tuples and takes log-linear time complexity. We show results in section~\ref{sss:fastText_exec_analy} with amortized batched inferencing time. It might seem that fine-grained parallel quicksort implementation can reduce individual inference time for datasets with a very large number of training instances and unique labels. Though, they scale well for large data sizes, achieving real-time latencies with such fine-grained approaches isn't feasible. Instead, reducing the instances retrieved in $I$ reduces the computational cost. This is because $I$ can be quite large, especially for datasets with a large number of labels and training points. This is because certain high-frequency words can be associated with a large number of instances, thus graph $G_{WI}$ can get quite large. This is mitigated by modifying lines 3-5 in algorithm~\ref{alg:graphite_cluster} to enable intrinsic SIMD vectorization by the compiler. But the size of $I$ input to line 6 is still large. So, during the de-duplication and counting in function $SDC$ in line 6 we avoid log-linear time complexity by using a count array to store the occurrence count of each instance while only performing linear operations. The sizes of $L$ and $R$ aren't large due to the cut-off for a pre-determined number of predictions at line 6.



\section{Experimentation and Results}
\label{ss:graphite_results}

 \subsection{Setup and Preprocessing}
\label{sss:graphite_setup}

Graphite is implemented for multi-core systems and don't require a GPU. The inference part is implemented in C++ ($\geq$ g++-9.3.0) using OpenMP threading with Python wrappers using \textit{pybind11}. We first compare our model with fastText \cite{fastText1,fastText2} as mentioned in section~\ref{sec:rel_work}. The training was done with the best optimal hyper-parameters searched using fastText's \textit{Automatic hyper-parameter optimization} with validation set. There was additional configuration tuning done only on the eBay datasets, such as setting the minimum frequency of words and labels, and so on. We can't discuss these details due to proprietary usage. The fastText model is based on version 0.9.2. \footnote{We used a system with 4 Intel Xeon Gold 6230 CPUs with 20@2.10GHz cores, 500 GB of RAM, and 2 Nvidia Tesla V100-32GB GPUs.}


We show comprehensive analysis with \emph{Astec}~\cite{dahiya2021deepxml} in the DeepXML framework. For each dataset, we use the configuration provided for the datasets in~\cite{Bhatia16} based on similar size label space. During the training, we could only choose the \emph{Label Clustering} configuration as other implementations weren't provided in the source code mentioned in ~\cite{dahiya2021deepxml}. We also test the feasibility of AttentionXML~\cite{you2019attentionxml} on the  \emph{Very Large} categories for our eBay datasets. 

\subsection{Notations and Metrics}
\label{sss:graphite_not_met}
We introduce some notations and metrics that are used for the analysis in next sections while borrowing terms from section~\ref{ss:notations}. A test set $S(A,B)$ contains item's title words or instances ($A$) and keyphrases or labels ($B$) as lists. For each sample $i$, we denote the corresponding title word list as $a_i\in A$ and keyphrases as $b_i\in B$, $A,B\in S$. The $a_i$'s are used as input and the $b_i$'s are used as \emph{ground truths} for the inference method described in the~\ref{ss:graphite_inf} to obtain the predicted labels $p_i$. For each sample $i$, we number the relevant predictions as $Relevance(i,k)=|b_i\cap p_i(k)|$, where $p_i(k)$ are the top $k$ predicted labels. For comparison we define the metrics $Precision@k=\frac{1}{|A|}\sum_{i=1}^{|A|}\frac{Relevance(i,k)}{|p_i(k)|}$, $Recall@k=\frac{1}{|A|}\sum_{i=1}^{|A|}\frac{Relevance(i,k)}{|b_i|}$. We find that in all the datasets, a large number of items are only associated with just one keyphrase. So, the Precision@k wouldn't correctly account for the items that have different ground truth counts. Hence, we devised a metric called \emph{Average Variable Precision~(AVP)} which quantifies what fraction of ground truth on an average does the model accurately predicts. It is defined as 
$AVP=\frac{1}{|A|}\sum_{i=1}^{|A|}\frac{Relevance(i,|b_i|)}{|b_i|}$. For AVP calculation we place a limit of 10 ground truths, so for test data points with $>10$ ground truths, only 10 predictions for each model are compared. 

\subsection{Performance on eBay Datasets}
\label{ss:eBay_results}
In this section, we show results on the proprietary datasets from eBay. We first describe the eBay datasets in subsection~\ref{sss:eBay_datasets}. The subsequent subsections compare Graphite with fastText and Astec on all the categories shown in Table~\ref{tab:category_id_grouping}. Due to proprietary constraints, we can't show absolute scores for the metrics described in section~\ref{sss:graphite_not_met}. Instead, we report all scores relative to fastText's performance which acts as a baseline.

\subsubsection{eBay Datasets}
\label{sss:eBay_datasets}
The datasets were generated from eBay engines' search logs. From among those items shown to a buyer, specifically the items that are clicked by the buyers are used. A buyer might input different queries/keyphrases in a session and click on some of the displayed items. Different buyers across various sessions who input the same keyphrase will be shown similar items which they eventually click on. Such occurrences of a keyphrase and the clicked item are considered as one sample keyphrase-item combination if they co-occur regularly. Similarly all such keyphrase-item associations are aggregated into a dataset if they co-occur for a sufficiently long period. It is intuitive to see how the keyphrases can be related to the items and also to the words in the titles. 

\begin{table}[h]
\scriptsize
\resizebox{\columnwidth}{!}{%
\begin{tabular}{@{}cc|cc|cc@{}}
\toprule
 &
   &
  \multicolumn{2}{c|}{\textbf{\# Train (in Millions)}} &
  \multicolumn{2}{c}{\textbf{\# Labels (in Millions)}} \\ \cmidrule(l){3-6} 
\multirow{-2}{*}{\textbf{Category ID}} &
  \multirow{-2}{*}{\textbf{Group}} &
  \multicolumn{1}{c}{\textbf{min}} &
  \textbf{max} &
  \multicolumn{1}{c}{\textbf{min}} &
  \textbf{max} \\ \midrule
\multicolumn{1}{c|}{\begin{tabular}[c]{@{}c@{}}CAT\_1-8\end{tabular}} &
  Very Large &
  \multicolumn{1}{c}{3.7} &
  25 &
  \multicolumn{1}{c}{2} &
  7 \\ \midrule
\multicolumn{1}{c|}{\begin{tabular}[c]{@{}c@{}}CAT\_9-16\end{tabular}} &
  Large &
  \multicolumn{1}{c}{2.4} &
  7.4 &
  \multicolumn{1}{c}{0.5} &
  1.7 \\ \midrule
\multicolumn{1}{c|}{\begin{tabular}[c]{@{}c@{}}CAT\_17-27\end{tabular}} &
  Medium &
  \multicolumn{1}{c}{0.8} &
  2.7 &
  \multicolumn{1}{c}{0.2} &
  0.6 \\ \midrule
\multicolumn{1}{c|}{\begin{tabular}[c]{@{}c@{}}CAT\_28-40\end{tabular}} &
  Small &
  \multicolumn{1}{c}{0.003} &
  0.6 &
  \multicolumn{1}{c}{0.003} &
  0.2 \\ \bottomrule
\end{tabular}%
}
\vspace{1mm}
\caption{Category ID and Grouping.}
\label{tab:category_id_grouping}
\end{table}

The datasets consist of training/validation/test sets per top level product category in eBay. We also grouped the categories into \emph{Very Large}, \emph{Large}, \emph{Medium}, and \emph{Small} based on the number of training points and a number of labels in each category. The anonymized categories (40) which are numbered in the non-increasing order of their training size and their groups are mentioned in Table~\ref{tab:category_id_grouping}. The table also shows the range of the number of training data points and labels for each group. The training set contains the bulk of the historical data, while validation and test sets for tuning and testing are limited to a few thousand data points.

\begin{figure*}[h]
\centering
\scriptsize
\includegraphics[width = 0.8\linewidth]{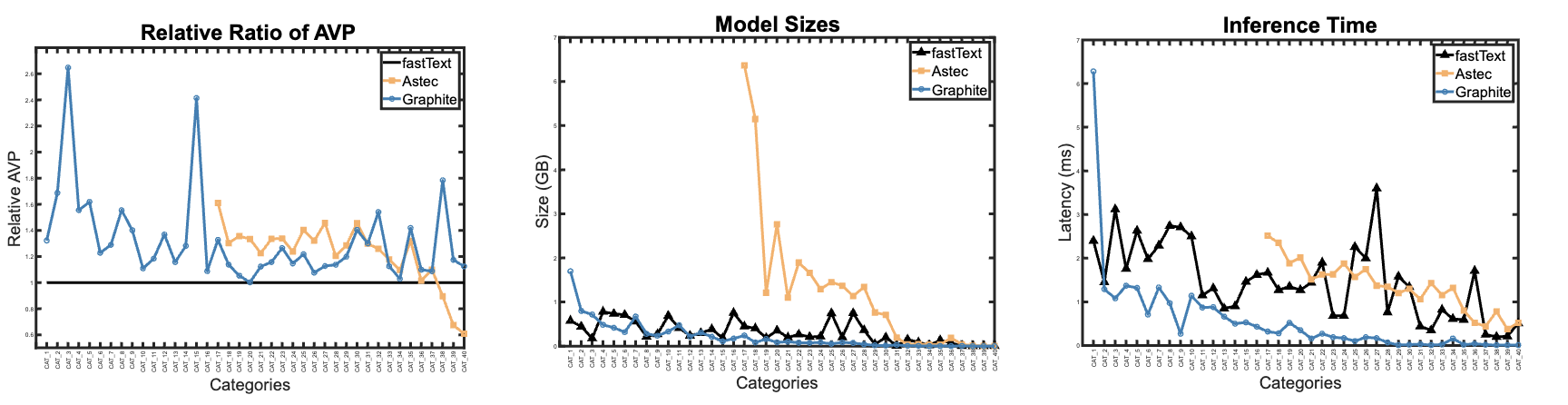}
\caption{Comparison of AVP, Model size and Inference time ratios of fastText, Astec and Graphite}
\label{fig:Graphite_model_perf}
\par \medskip
\end{figure*}

\subsubsection{Prediction Performance}
\label{sss:eBay_pred_results}
For each category, the top $10$ predictions are used for comparison on the test set. fastText's baseline is set to 1 for all categories while Graphite's and Astec's performance is shown relative to it. We show the comparison of Relative Ratios of Precision@1/5/10 for the models in figure~\ref{fig:graphite_scores_precision} for all categories. The categories in the x-axis follow the order of \emph{Very Large} to \emph{Small} group as in table~\ref{tab:category_id_grouping}. Graphite shows better performance than fastText for all the categories across all scores, especially with an average improvement of $122\%$ in Precision@1 with up to 210\% improvement for CAT\_3. For all precision scores, the gap in the precision score is generally higher for categories with larger label spaces, while precision score gap is typically lower for categories with smaller label spaces. The Recall scores in figure~\ref{fig:graphite_scores_recall} also show that Graphite has better performance w.r.t fastText, especially with Recall@10 scores gaining an average improvement of $140\%$ with highest being $270\%$. In Precision/Recall@$k$ with $k>1$, the gap between Graphite and fastText is higher than the trend of $k=1$. This is due to fastText's label space limitation which excludes tail keyphrases that Graphite is easily able to capture. The \emph{Average Variable Precision} (AVP) scores in figure~\ref{fig:Graphite_model_perf} show average performance akin to a precision score. Graphite performs better than fastText with an average increase of $0.085$ with up to $0.207$ with respect to fastText. The chart shows that Graphite on average is better able to recommend at least one of the ground truths for the items.


On the other hand, Astec fails to execute on categories CAT\_1-16 in figures~\ref{fig:graphite_scores} and~\ref{fig:Graphite_model_perf}. For instance, it tries to allocate more than 1 TB of RAM for CAT\_11 which was larger than the system's memory resources. The requirements are much larger for other \emph{Large/Very Large} categories. Astec's clustering process doesn't scale well for a large number of training points. Similarly, DNNs such as AttentionXML also fail on these large datasets while allocating large GPU memory. More details regarding their executions are discussed in section~\ref{sss:astec_hugedata}.  

Astec's performance is higher than fastText's for medium and small size categories with average P@1 and R@10 improvement of $115\%$ and $124\%$ respectively. However, it underperforms on very small size categories, especially on CAT\_39 and CAT\_40, across all precision, recall and AVP scores due to relatively smaller number of training points than the number of labels. 
On the categories that Astec executes on, Graphite is comparable to Astec, with an average gap of $\pm 0.011$ between the relative performances of both the models. In contrast to Astec, Graphite generally shows consistently better performance across all data sizes, except for a few \textit{Large} and \textit{Medium} categories.
\vspace{-2mm}
\subsubsection{Execution performance}
\label{sss:fastText_exec_analy}
In this section, we show comparisons based on model size, inference time, and training time. If we look at the top 3 categories (CAT\_1-3) with the largest label space, Graphite's trained model occupies $2-3\times$ fastText's storage on disk as shown in the middle chart of figure~\ref{fig:Graphite_model_perf}. For all the other categories Graphite's model size is comparable or occupy lower space on disk than fastText. Infact, if we sum up the model sizes of all categories, Graphite occupies $30\%$ less space than fastText. As Graphite's storage is a linear function of both the number of training data points and unique labels, the models of larger categories occupy more space than the smaller ones. This can be tuned by eliminating rarely occurring keyphrases along with those training items that don't have any keyphrases left after the removal. This might not significantly reduce the performance when retrieving the tail keyphrases. On the other hand, as expected Astec's models are extremely large, occupying \emph{1.2 GB} on average per category.
\begin{figure}[]
    \centering
    \includegraphics[width=0.6\linewidth]{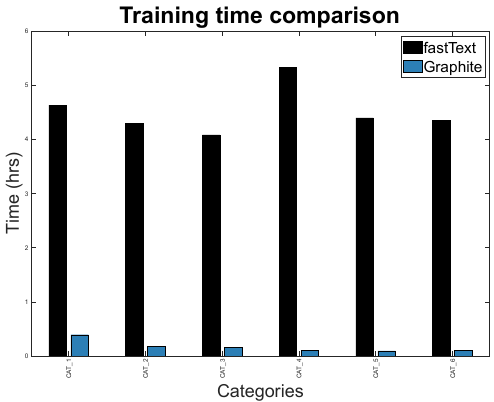}
    \vspace{1mm}
    \caption{Training times of fastText and Graphite for the top 6 categories.}
    
    \label{fig:graphite_training_time}
    \vspace{3mm}
\end{figure}

We compute the inference time per test data point by amortizing the time taken for inferencing the entire test dataset. In Graphite, we employ a coarse-grained multi-threading technique that executes individual inferences on separate threads whereas fastText is single-threaded. We launch a variable number of OMP threads (max 26) and set the number of predictions to 10. The right-most chart of figure~\ref{fig:Graphite_model_perf} shows the per-sample inference times for fastText, Astec and Graphite on all the categories. For all except the largest category,  the inference times of both models are comparable, with a geometric mean speedup of $7\times$ of Graphite over fastText. Graphite is much faster than Astec with an average speedup of $20\times$. Astec's inference was done on GPU due to absence of any CPU based code. Graphite's inference time on Small datasets group is extremely small with upto $90\times$ speed up over both fastText and Astec. Since, CAT\_1 is the largest category
by label space and number of items, Graphite results in a longer
inference time due to the reasons discussed in section~\ref{sss:illustration_implementation}. We mitigate this practically by using a relatively smaller training set
for CAT\_1 corresponding to a shorter historical duration. Although this results in a smaller number of CAT\_1 items, it still has sufficient number of items resulting in minimal impact on the recommendation.


Figure~\ref{fig:graphite_training_time} shows the training times required to generate a model for both fastText and Graphite. We only compare using categories from the \emph{Very Large} group from Table~\ref{tab:category_id_grouping}, due to their much larger training times. The \emph{autotune} duration of fastText was set to a certain number of hours to choose the best hyperparameter for optimal model size and precision/recall scores. In some cases, getting an optimal score required autotuning for $\sim 10$ hours. The average training time of categories on which Astec executes was $1.8$ hours with up to $7$ hours for some categories. This excludes the time to compute the surrogate clustering stage which runs into hours. The categories that Astec fails to execute on, would require much larger epochs resulting in training time running into days. In contrast, Graphite takes only a few minutes to generate a model and thus, enables frequent model refresh and efficient model management.

\vspace{-2mm}
\subsubsection{Shortcomings of DNN models on Huge Datasets}\label{sss:astec_hugedata}
To discuss Astec's failure to execute on \textit{Very Large} Categories, we first dive into more detail on its \emph{surrogate} task which it's pivotal stage. Astec's surrogate stage performs two steps in conjunction, Label selection and Intermediate representation training. As described in section~\ref{sec:rel_work}, the selection is done using clustering which requires generating a representation per label. This representation is generated from those training data points that each label is associated with. To perform the clustering, Astec tries to allocate memory in RAM as a function of the number of training points and labels. Thus, it fails for all the categories in \emph{Very Large} and \emph{Large} groups due to the infeasibility of the allocation for their sizes as mentioned in table~\ref{tab:category_id_grouping}. 

AttentionXML also doesn't run for the \emph{Very Large} categories. It encounters limitations due to its substantial CUDA memory requirements for conducting the forward pass and computing gradients during its multi-label attention phase. Even for category CAT\_8, AttentionXML requires more GPU memory than two Tesla V100 ($>$64GB) to train the model. This requirement would be even larger for categories CAT\_1-7, and theoretically for CAT\_1, the requirement could exceed the specifications of commercial GPUs even when the hidden layer's size is restricted to 64.
\vspace{-2mm}

\subsubsection{AI Evaluations and Case Study}
\label{sss:ai_human_evaluations}
The keywords predicted were evaluated for a subset of randomly sampled 100 items using GPT-4 \cite{openai2023chatgpt} evaluation as a proxy for human evaluation. The responses were \textit{yes/no} answers to relevance between the title and keywords. Table~\ref{tab:ai_evaluations} shows the percentage of item-keyword pairs in comparison to fastText that were evaluated as yes for one \textit{representative category} from each group in table~\ref{tab:category_id_grouping} including the ground truth keywords for the subset of items. \footnote{All the GPT-4 numbers for the ground-truth are greater than 90\% showing high-degree of alignment with positive buyer judgement.} \textit{Graphite shows superior alignment with GPT-4 evaluation.}

\begin{table}[h]
\centering
\begin{tabular}{@{}c|c|cccc@{}}

\toprule
Group & Category & fastText & Astec & Graphite &  Ground Truth \\ \midrule
\multicolumn{1}{c|}{Very Large} & \multicolumn{1}{c|}{CAT\_4}  & 30.6 & -    & \textbf{61.2} & 92.9 \\
\multicolumn{1}{c|}{Large} & \multicolumn{1}{c|}{CAT\_11} & 44.6 & -    & \textbf{72.8} & 90.9 \\
\multicolumn{1}{c|}{Medium} & \multicolumn{1}{c|}{CAT\_18} & 61.2 & 76.6 & \textbf{78.0} & 91.9 \\
\multicolumn{1}{c|}{Small} & \multicolumn{1}{c|}{CAT\_28}                      & 50.4 & 72.8 & \textbf{75.5} & 91.1 \\ \bottomrule
\end{tabular}%
\vspace{2mm}
\caption{Percentage of relevant keywords from AI evaluations of 10 predictions for 100 items from fastText, Astec and Graphite.}
\label{tab:ai_evaluations}
\end{table}

A case study on the item titled \textit{``treated wooden deckboards''} demonstrates the effectiveness of each model's top recommendation. Graphite predicts \textit{deckboards} which is the same as the ground truth and found relevant by GPT4. In contrast, fastText and Astec predict \textit{moraea} and \textit{wooden stakes} respectively, both being annotated irrelevant. For another item \textit{``bird bath stone water basin garden decoration food bowl frost resistant''}, Graphite predicts \textit{stone bird bath} while both fastText and Astec predict the irrelevant keyphrase \textit{floor bird bath}. In both examples, Graphite's predictions have similar words to the item's title, thus our model is better able to find the nuances between different labels recommended for an input. Whereas, fastText and Astec embeddings yield irrelevant results due to imperfect learning. 




\vspace{-2mm}
\subsubsection{Impact}
\label{sss:impacts}
 Graphite was deployed as part of eBay's seller side keyphrase-recommendation service. This was possible due to its scalability and lightweight nature. A differential pre-post analysis was performed \textemdash which showed that Graphite was able to increase the eBay platform's coverage by recommending 6\% more unique keyphrases and 17\% more unique item-keyphrase pairs (keyphrase coverage). The \textit{acceptance rate} --- the fraction of keyphrases out of the recommended keyphrases accepted by the sellers to place bids on --- for Graphite keyphrases was 3\% higher than fastText\footnote{All these numbers were calculated over a period of one month for over millions of keyphrases and thousands of sellers.}. The lift in acceptance rate reinforces the fact that \textit{Graphite keyphrases are better in terms of seller (human) judgement than fastText}, as sellers show a greater proclivity to bid on these keywords than fastText. We cannot disclose any more information due to proprietary and business constraints.

\subsection{Performance on Public Datasets}
\label{ss:graphite_otherDatasets}

Apart from the analysis of the eBay datasets, we also wanted to compare the performances of the models on other accessible datasets for the sake of reproducibility and absolute comparison. The applicable datasets should be search-based containing queries/keyphrases that bear similarities to the input text. The label text is essential for the comparison as our model uses word matches in the label. The standard datasets in the Extreme Multi-Label Short Text classification space~\cite{Bhatia16} are derived from real-world applications, ranging from item-to-item recommendation (\emph{AmazonTitles-670K},  \emph{AmazonTitles-3M}, etc.), to text-category tagging (\emph{AmazonCat-13k}, \emph{Wikipedia-500K},  etc.). However, for the purpose of keyphrase recommendation, these datasets were not suited for our analysis.
 
For our analysis, we hinged on publicly available keyphrase recommendation datasets available in~\cite{bworld}. Out of these datasets, we picked \emph{KPTimes}~\cite{gallina2019kptimes} and \emph{KP20k}~\cite{meng2017deep} which are the largest datasets to compare for scalability. 
KPTimes has 102,357 unique labels associated with 259,923 training data points which are small enough for large models to handle. While KP20k has 680,117 uniques labels for 514,154 data points. 

Table ~\ref{tab:kptimes_performance} shows the absolute performances of fastText, Astec and Graphite on KPTimes and KP20k. The Precision@5 and AVP scores of Astec for KPTimes were better than all the models while Astec couldn't provide any meaningful predictions for KP20k. For other scores, all model's performances were the same and Graphite's performance stood out in execution metrics. \footnote{fastText's per inference time on the datasets in table~\ref{tab:kptimes_performance} is relatively higher as the model's autotuning was carried out \textit{only} using its \emph{autotune} parameter. The tuning of parameters such as minCount and minCountLabel reduces the sizes of the token and label spaces, thus reducing inference time.}


\begin{table}[h]
\centering
\resizebox{\columnwidth}{!}{%
\begin{tabular}{@{}lcccccc@{}}
\toprule
\multirow{2}{*}{Metrics}                                   & \multicolumn{3}{c}{KPTimes}                              & \multicolumn{3}{c}{KP20k}            \\ \cmidrule(l){2-7} 
                                                           & fastText & Astec         & \multicolumn{1}{c|}{Graphite} & fastText & Astec & Graphite \\ \midrule
\multicolumn{1}{l|}{P@1}                                   & 0.39     & 0.41          & \multicolumn{1}{c|}{\textbf{0.41}}     & 0.10     & -     & \textbf{0.28}     \\
\multicolumn{1}{l|}{P@5}                                   & 0.21     & \textbf{0.25} & \multicolumn{1}{c|}{0.21}     & 0.05     & -     & \textbf{0.12}     \\
\multicolumn{1}{l|}{P@10}                                  & 0.14     & 0.14          & \multicolumn{1}{c|}{\textbf{0.14}}     & 0.04     & -     & \textbf{0.08}     \\
\multicolumn{1}{l|}{R@1}                                   & 0.08     & 0.08          & \multicolumn{1}{c|}{\textbf{0.08}}     & 0.02     & -     & \textbf{0.05}     \\
\multicolumn{1}{l|}{R@5}                                   & 0.21     & 0.21          & \multicolumn{1}{c|}{\textbf{0.21}}     & 0.05     & -     & \textbf{0.12}     \\
\multicolumn{1}{l|}{R@10}                                  & 0.27     & 0.27          & \multicolumn{1}{c|}{\textbf{0.27}}     & 0.07     &       & \textbf{0.16}     \\
\multicolumn{1}{l|}{AVP}                                   & 0.22     & \textbf{0.25} & \multicolumn{1}{c|}{0.22}     & 0.05     & -     & \textbf{0.13}     \\
\multicolumn{1}{l|}{\begin{tabular}[c]{@{}l@{}}Inference \\ Time (ms)\end{tabular}} & 6.2 & 0.32 & \multicolumn{1}{c|}{\textbf{0.03}} & 67.7 & 0.36 & \textbf{0.05}  \\
\multicolumn{1}{l|}{\begin{tabular}[c]{@{}l@{}}Training \\ Time\end{tabular}}       & 4h  & 0.5h & \multicolumn{1}{c|}{\textbf{7.5s}} & 4h   & 1.8h & \textbf{21.2s} \\
\multicolumn{1}{l|}{\begin{tabular}[c]{@{}l@{}}Model \\ Size (MB)\end{tabular}}       & 47       & 595           & \multicolumn{1}{c|}{\textbf{16}}                   & 324      & 299   & \textbf{50} \\ \bottomrule
\end{tabular}%
}
\vspace{2mm}
\caption{Performance of fastText, Astec and Graphite on KPTimes and KP20k dataset.}
\label{tab:kptimes_performance}
\vspace{-3mm}
\end{table}

\section{Conclusions and Future Work}
We present a \textit{simple}, \textit{transparent}, and \textit{lightweight} graph model that can classify an extremely large number of labels in real-time. We show comparisons with baseline models like fastText and with a state-of-the-art model like Astec using real-world datasets provided by eBay. We find that fastText performs reasonably well on the extreme classification task due to its linear neural network architecture. Graphite's performance is better than that of fastText based on the Precision, Recall, and AVP scores for the 40 eBay categories that we analyzed. Both our model and fastText can handle larger datasets. Although Astec's performance is comparable to Graphite, it fails to execute on \emph{Very Large} and \emph{Large} categories due to its usage of \textit{centroid} method in the surrogate task stage. Graphite has lower training time than fastText with its inference time comparable to fastText. In the future, we aim to apply Graphite to a variety of other classification tasks especially where the label text shares words with the input texts. We find that the clustering phase of our model is crucial to its performance, which can be further improved with better and light-weighted clustering algorithms. The execution and storage cost of the large category in the eBay dataset can be further mitigated by developing distillation techniques that reduce the number of training points without compromising any significant information. Graphite's bipartite graph can also include weights indicating relational propensity among the words, instances, and labels.






\newpage
\bibliography{refs}
\newpage
\appendix
\section{Appendix}

\subsection{Ablation Studies}
\label{ss:ablation_studies}

\subsubsection{Word Match Ratio}
To compute the \emph{Word Match Ratio}, the constituents of each label are stored in CSR format with each label ID mapped to the word IDs that constitute the label. Thus, the time complexity of the intersection for the substring comparison is $O(|t|)$. 

One drawback of the tokenized word comparison is that it doesn't account for alternate words and different word forms. This can be mitigated by computing the sub-words of each word token and then matching sub-words of both lists of words. However, in our experiments (which is outside the scope of this work) it only has a slight improvement in the performance of our method while increasing the run time to $O(|t|\cdot x)$ where $x$ is the maximum number of possible sub-words. It requires additional storage of the mapping of word IDs to the sub-word IDs. We suspect that there aren't a lot abnormal word forms in our datasets to significantly affect the recommendations. We leave further exploration on this topic for future research.

\subsubsection{Graphite Alternate Ranking}
\label{sss:graphite_alt_ranking}
Prior to finalizing the ranking strategy discussed in section~\ref{sss:graphite_ranking}, we investigated various other strategies on the eBay datasets. We looked at performances of two alternate \emph{Word Match Ratios}; Jaccard Similarity and the ratio of count of common words to number of missing words in the label. The drop in AVP scores for each of the above ratios with respect to the word match ratio ranged $0.025-0.050$ and $0.009-0.036$, respectively, across all eBay categories. The latter ratio ranks relatively large length labels higher than shorter labels. In context of e-commerce listings, this ratio prefers more specific keywords like ``iPhone 14 pro max'' to generic keywords like ``iPhone''. Though its precision is lower, such keywords are advantageous to the sellers. 

Further analysis was done by changing the order of ranking corresponding to line 15 of Algorithm~\ref{alg:graphite_cluster}. The Word Match Ratio was positioned before the item/training point similarity score. Thus, sorting of labels was first done by $WMR(\cdot,\cdot)$ then the ties were broken by the similarity score. The current ordering still performs better than above mentioned ordering with an average AVP gap ranging $0.012-0.95$ across all categories. This shows that the Word Match ratio is an important metric for distinguishing among the labels while similarity score gives us an edge. If we merge the top 2 clusters of labels identified by the similarity score, we get atmost $0.020$ drop in AVP compared to the current metric. This allows for more training items to be matched to the test item in the top cluster.

\subsubsection{Synthetic Data}
\label{sss:graphite_synth_data}
We devised a dataset to further understand the role that Graphite's item similarity score plays. The dataset generated had training, test, validation and label sizes as 500,000, 10,000, 10,000 and 200,000 respectively. Each sample in the dataset has a title with 10 words associated with exactly 10 labels. Each sample in the test and validation set has exactly one equivalent training sample to match with. All the words in the test sample will match with the corresponding training sample and the labels of such sample are the ground truths for the test/validation sample. Graphite has a perfect score of \num{1.0} for P@1, P@5 and AVP. While fastText achieves \num{0.037}, \num{0.017} and \num{0.012} for the same. Astec lags behind fastText with scores \num{0.016}, \num{0.013} and \num{0.012} for P@1, P@5 and Astec respectively. Graphite employs a similar item recommendation strategy which given a test item, first finds the most similar training item and then recommends its labels. This helps in achieving a perfect score which other models fail to do due to imperfect learning.

\subsection{Interpretability}
\label{sss:graphite_interpretability}
Commercial operations often require a model to be interpretable to understand the reasoning behind the predictions and decisions made by the model. Neural Network model generally require vectorizing the input text which often leads to obscured understanding of each token's contribution to the decisions. Various gradient optimization techniques also conceal the contribution of each input to the weight matrix. Moreover, explainable methods such as LIME and SHAP provide posterior explanations treating a DNN model as a black-box. 

Let's consider the following scenario: Given a test input and a model's predictions, if the recommendations are considered to be problematic, it's essential to identify the probable causes. For example, if the keywords for "iPhone" phone models are often recommended for products that are "iPhone" accessories, it can be misleading to buyers. In such cases, it is necessary to understand the decision path of the model and also spot the erroneous training data points. Utilizing explainable methods are infeasible since they'll require iterating over the entire training dataset to evaluate each data point's contribution. Graphite's Algorithm~\ref{alg:graphite_cluster} also has the ability to store the item IDs and similarity score after line 6. Since only top 2 clusters are sufficient, using the similarity scores and item IDs we'll have limited number of items that are considered more relevant to the test item. This significantly narrows down the search for potential data points and the candidate labels that are problematic.

\subsection{AI Evaluation Prompts}
\label{sec:prompts}

Below is the prompt for GPT-4 as a keyphrase relevance evaluator, with templated values to be replaced by the item title and the model generated keyphrases.

\noindent\texttt{\small{Given an item with title: \{\textrm{title}\},
determine whether below keywords are relevant by giving yes or no one by one as answer: \\
 1. \{\textrm{kw1}\};\\
 2. \{\textrm{kw2}\};\\
 3. \{\textrm{kw3}\};\\
 4. \{\textrm{kw4}\};\\
 5. \{\textrm{kw5}\}.\\
 \#\#\#Response:
 }}
\end{document}